\documentclass[11pt,twoside]{article}
\usepackage{asp2010}
\resetcounters
\aspvolume{455}
\aspcpryear{2012}
\aspvoltitle{4$^{\rm th}$ {\em Hinode} Science Meeting: Unsolved 
Problems and Recent Insights}
\aspvolauthor{L.R.~Bellot Rubio, F.~Reale, and M.~Carlsson, eds.}

\begin{document}

\title{Supersonic Magnetic Flows in the Quiet Sun Observed with \textsc{Sunrise}/IMaX}
\author{J.~M.~Borrero,$^1$ V.~Mart{\'\i}nez Pillet,$^2$ 
R.~Schlichenmaier,$^1$ W.~Schmidt,$^1$ T.~Berkefeld,$^1$
S.~K.~Solanki,$^{3,4}$ J.~A.~Bonet,$^2$ J.~C.~del Toro Iniesta,$^5$
V.~Domingo,$^6$ P.~Barthol,$^3$ and A.~Gandorfer$^3$
\affil{$^1$Kiepenheuer-Institut f\"ur Sonnenphysik, 
Sch\"oneckstr.~6, D-79104, Freiburg, Germany\\
$^2$Instituto de Astrof{\'\i}sica de Canarias, V{\'\i}a L\'actea
s/n, 38200 La Laguna, Spain\\ $^3$Max Planck Institut f\"ur
Sonnensystemforschung, D-37191, Katlenburg-Lindau, Germany\\
$^4$School of Space Research, Kyung Hee University, Yongin, Gyeongg
446-701, Republic of Korea\\ $^5$Instituto de Astrof{\'\i}sica de
Andaluc{\'\i}a (CSIC), Apdo.~de Correos 3004, Granada, Spain\\
$^6$Image Processing Laboratory, University of Valencia, 
P.O.~Box 22085, E-46980 Paterna, Valencia, Spain}}

\begin{abstract}
In this contribution we describe some recent observations of
high-speed magnetized flows in the quiet Sun granulation.  These
observations were carried out with the Imaging Magnetograph eXperiment
(IMaX) onboard the stratospheric balloon {\sc Sunrise}, and possess
an unprecedented spatial resolution and temporal cadence. These flows
were identified as highly shifted circular polarization (Stokes $V$)
signals. We estimate the LOS velocity responsible for these shifts to
be larger than 6~km~s$^{-1}$, and therefore we refer to them as {\it
supersonic magnetic flows}. The average lifetime of the detected
events is 81.3~s and they occupy an average area of about
23\,000~km$^2$. Most of the events occur within granular cells and
correspond therefore to upflows. However some others occur in
intergranular lanes or bear no clear relation to the convective
velocity pattern. We analyze a number of representative examples and
discuss them in terms of magnetic loops, reconnection events, and
convective collapse.
\end{abstract}

\section{Introduction}

The interaction between convective motions and the magnetic field in
the quiet solar photosphere gives rise to a very rich variety of
physical processes: flux emergence/submergence, flux
reconnection/dissipation, convective collapse, thermal relaxation, and
so forth. Most of these processes occur at very small spatial and
short temporal scales, making their direct observations extremely
difficult.  Fortunately, the spatial resolution, temporal cadence and
polarimetric accuracy of modern instruments have improved to a point
where these observations are now becoming available, thus making it
possible to directly confront theory with observations. Undoubtedly,
the spectropolarimeter on-board of the {\em Hinode} satellite has
played a crucial role in this advancement, helping to uncover many of
these phenomena: emergence of magnetic loops
\citep{2007ApJ...666L.137C,2009ApJ...700.1391M,2009RAA.....9..921Z,2009A&A...495..607I,2010ApJ...714L..94M},
convective collapse in intergranular lanes
\citep{2008ApJ...677L.145N,2009A&A...504..583F}, horizontal supersonic velocities
\citep{2009ApJ...700..284B,2010MmSAI..81..751S}, supersonic downflows \citep{2007ASPC..369..113S},
horizontal internetwork fields
\citep{2009A&A...495..607I}, and many more.

A major step has also been achieved thanks to the stratospheric
balloon {\sc Sunrise}, which was launched from ESRANGE (Kiruna,
Sweden) in June 2009 \citep{2010ApJ...723L.127S,2011SoPh..268....1B}. During its
flight, one of the scientific instruments on-board of {\sc Sunrise},
the Imaging Magnetograph eXperiment \citep[IMaX;][]{2011SoPh..268...57M}
recorded spectropolarimetric data with an unprecedented spatial
resolution (0\farcs15) and temporal cadence (32 seconds). These data have
already been employed to reveal further details about the dynamics of
the magnetic field in the quiet Sun, such as vortex flows
\citep{2010ApJ...723L.139B}, vortex tubes \citep{2010ApJ...723L.180S}, dynamics of
horizontal internetwork fields \citep{2010ApJ...723L.149D}, and supersonic
magnetic upflows \citep{2010ApJ...723L.144B}. The latter paper uncovered the
existence of highly blueshifted polarization signals at the
center/edges of granular cells, that were interpreted as the signature
of supersonic magnetic upflows. Those authors also found evidence that
supports the idea of these flows being caused by magnetic
reconnection. However, only about 70\% of the detected events could be
ascribed to this scenario. In this contribution we investigate in more
detail some of the events that were found and present a detailed study
on whether they harbor supersonic velocities.

\section{Observations}

During {\sc Sunrise}'s arctic flight, the IMaX instrument recorded
polarimetric data (Stokes $I$, $Q$, $U$ and $V$) in five wavelength
positions across the \ion{Fe}{i}~5250.217~{\AA} (g$_{\rm eff}=3$)
spectral line. A full observing cycle was completed every 32
seconds. The spatial resolution of the observations has been estimated
to be $0\farcs15 \times 0\farcs18$ in a $46\arcsec \times 46\arcsec$
field of view. A number of factors contributed to IMaX's unprecedented
spatial resolution. First and foremost, at a height of 35 kilometers
most of Earth's atmospheric disturbances could be avoided. In
addition, a great pointing stability was achieved by the Correlation
tracker and Wavefront Sensor
\citep[CWS; ][]{2004SPIE.5489.1164S,2011SoPh..268..103B}. Finally, the excellent
optical quality of the telescope and instruments left few remaining
seeing effects and aberrations, which were subsequently corrected
employing a phase-diversity calibration of the point spread function
of the optical system.

The five tuning positions for IMaX dual-pass Fabry-Perot were
$\lambda-\lambda_0 = -80, -40$, $40, 80$, and 227 m{\AA}, with
$\lambda_0=5250.217$~{\AA} being the observed (solar) central
wavelength for the aforementioned neutral iron line. The last tuning
position, hereafter referred to as \emph{continuum}, was employed in
order to correct any residual cross-talk after the polarimetric
calibration was performed. Even after the removal of excess
cross-talk, a number of features could be detected in the 2D images of
the circular polarization in the continuum, $V_{\rm c}$ \citep[see
Fig.~1 in][]{2010ApJ...723L.144B}\footnote{In addition, movies of $V_{\rm c}$
can be found at
\url{ftp://ftp.kis.uni-freiburg.de/personal/borrero/sunrise/}}. These
features are spatially and temporally coherent, which points towards a
solar origin and rules out instrumental effects as their origin. The
value of the circular polarization in the continuum wavelength
($V_{\rm c}$) exceeds 1.5\% (in units of the quiet Sun continuum
intensity $I_{\rm qs}$). A total of 87 events were detected in the
combined 54.3 minutes of observations that we analyzed. The average
lifetime and area of these events are 81.3~s and $\approx$
23000~km$^2$, respectively.

As mentioned in \cite{2010ApJ...723L.144B}, about 70\% of these events are
related to the appearance of magnetic fields of opposite polarity in
their neighborhood. Whenever this happens, strong linear polarization
signals (revealing the existence of horizontal fields) are usually
seen in the pixels surrounding the event's location.

\begin{figure*}
\begin{center}
\includegraphics[width=12cm]{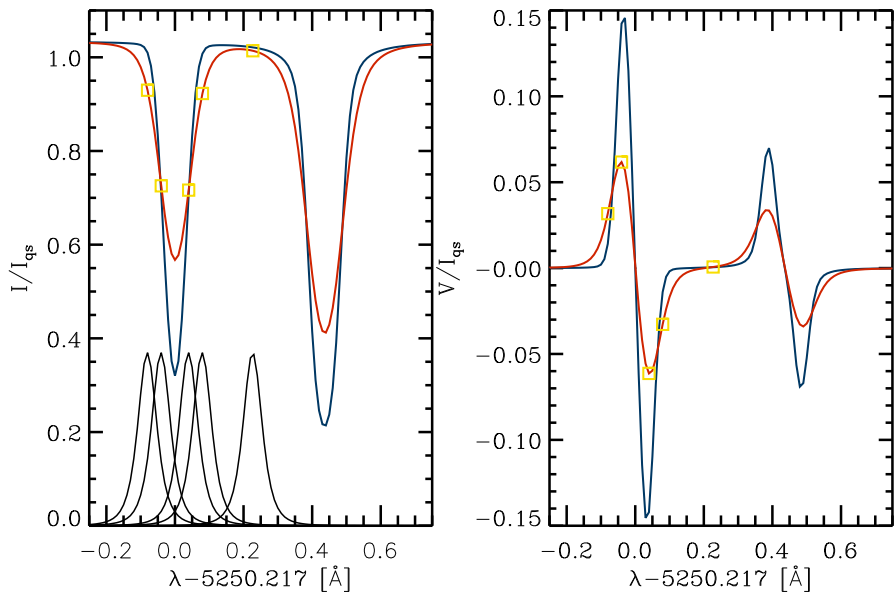}
\caption{Synthetic spectrum of \ion{Fe}{i}~5250.2 and 5250.6~{\AA} using
 the temperature stratification of the granular model from
 \cite{2002A&A...385.1056B}, but setting the velocities to zero and adding and
 homogeneous vertical magnetic field of 250~G (blue lines; Stokes $I$
 in the {\em left} panel and Stokes $V$ in the {\em right} one). Convolution of
 the blue profiles with IMaX's transmission profile (thin black curves
 at the bottom) yields the red curve. The yellow squares indicate the
 values measured by IMaX at each of the 5 wavelength positions.}
\end{center}
\end{figure*}

\section{Signature of Supersonic Velocities}

In order to produce polarization signatures 227 m{\AA} away from
$\lambda_0$, line-of-sight (LOS) velocities of up to 12~km~s$^{-1}$
might be needed. This value however does not take into account the
effects of the magnetic field, which shifts the $\sigma$-components in
Stokes $V$ away from $\lambda_0$. It also neglects the thermal and
Doppler width of the spectral line, as well as the additional
broadening induced by IMaX's filter profiles. In order to study these
effects, we have performed a numerical experiment in which we produce
synthetic Stokes profiles with varying LOS velocity. This is done with
the aim of finding the minimum velocity required to produce
$V_{\rm c}$ signals of about 1.5\% (in units of the quiet Sun continuum
intensity).

To this end, we have employed the synthesis module of SIR
\citep[Stokes Inversion based on Response Functions;][]{1992ApJ...398..375R} to
obtain synthetic Stokes $I$ and $V$ profiles for \ion{Fe}{i}~5250.217~{\AA}
and \ion{Fe}{i}~5250.653~{\AA}. The atmospheric model used was the granular
model from \cite{2002A&A...385.1056B}, but adding a vertical magnetic field
with the following characteristics: $B=250$ G, $\gamma=0^{\circ}$
(strength and inclination of the magnetic field vector,
respectively). Figure 1 shows the results for Stokes $I$ and $V$ in
blue color. These profiles are then convolved with IMaX's transmission
profiles (see Mart{\'\i}nez Pillet et al.  2011), which are shown in
black. The result of the convolution are the broadened red solid
profiles.  Finally, the yellow squares show the measurements taken by
IMaX, whose tuning positions were given in Sect.~2.  $V_{\rm c}$
corresponds to the rightmost yellow square in the right panel of
Fig.~1. For this particular figure we employed a LOS velocity of 0 km
s$^{-1}$ and therefore $V_{\rm c} \sim 0$. However it is clear that,
for very large blueshifted or redshifted velocities, a signal due to
\ion{Fe}{i}~5250.653~{\AA} or \ion{Fe}{i}~5250.217~{\AA} respectively, should appear
in $V_{\rm c}$.

\begin{figure*}[t]
\vspace*{-.5em}
\begin{center}
\hspace*{-2em}
\includegraphics[width=10.6cm]{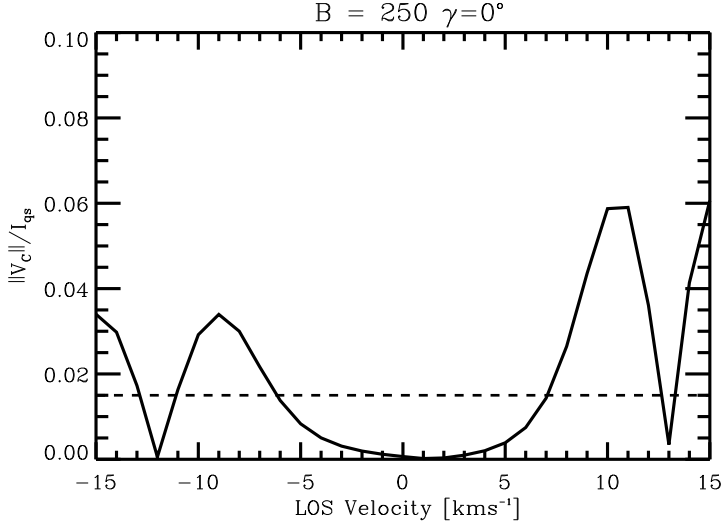}
\vspace*{-.5em}
\caption{Absolute value of the circular polarization observed by 
IMaX in $\lambda_{\rm c}$ as a function of the LOS velocity, for a granular
model with a vertical (inclination of the magnetic field with respect
to the observer $\gamma=0^{\circ}$) magnetic field of $B=250$~G.}
\end{center}
\end{figure*}

Figure 2 shows precisely this: the amount of $V_{\rm c}$-signal
detected as a function of the LOS velocity employed in the
synthesis. The horizontal dashed line corresponds to the lower
threshold for the detection of these events: $|V_{\rm c}|/I_{\rm
qs} > 0.015$. This figure indicates that, once all additional
broadening effects have been considered, LOS velocities larger than 6
km s$^{-1}$ are needed to produce $V_{\rm c}$-signals close to the
observed values. Positive (redshift or downflow) and negative
(blueshift or upflow) velocities are both possible, since the
continuum point lies in between the two spectral lines \ion{Fe}{i}~5250.217
and 5250.653~{\AA}.

Note that this experiment was performed with a vertical magnetic field
strength of 250 G. Increasing the value of longitudinal component of
the magnetic field will certainly decrease the absolute value of the
velocity required to produce sufficient signal in $V_{\rm
c}$. However, from the measurements of the circular polarization in
these events we have determined 250 G to be an upper limit. Therefore
we conclude that, in order to produce $| V_{\rm c} | / I_{\rm qs} >
0.015$, LOS velocities equal to or larger than 6~km~s$^{-1}$ in
absolute value are required. Considering that the total velocity is
larger or equal than the LOS velocity, and that the speed of sound in
the solar photosphere is about 6~km~s$^{-1}$, we can ascribe the
observed signal as being produced by supersonic and magnetized
(because the polarization signal is shifted) flows. For this reason we
will hereafter refer to these events as \emph{supersonic}.

\section{Statistics and Interpretation}

\begin{table*}[t]
\centering
\tabcolsep 2.3em
\caption{Event classification in terms of the velocity fields 
around the event location.  See text for further details.}
\begin{tabular}{lcc}
\noalign{\smallskip}
\tableline
\noalign{\smallskip}
 Type & \# Events &\% from total \\
\noalign{\smallskip}
\tableline
\noalign{\smallskip} 
Cell center & 44 & 51 \\
Cell edge & 17 & 19 \\
Intergranule & 5 & 6 \\
Evolving granulation & 6 & 7 \\
Exotic event & 7 & 8 \\
Unclear & 8 & 9 \\
\hline
\end{tabular}
\end{table*}

We have classified the observed supersonic events in two different
ways. The first classification was done attending to the number of
patches of enhanced $V_{\rm c}$ that appear. If one single patch of
$V_{\rm c}/I_{\rm qs} > 0.015$ is seen we refer to it as
\emph{single}. If two patches of enhanced $V_{\rm c}$ appear next to
each other (within 2\arcsec\/), we refer to this event as \emph{twin}
(if the two patches have the same sign in $V_{\rm c}$) or
\emph{double} (opposite sign). The most common type of event is
the single one (78 cases out of 87, or 82\%), whereas double
events are rare (14 cases, 15\%) and twin very rare (3 cases, 3\%).

The second classification has been done in terms of the LOS velocity
at the location of the event. Results are presented in Table~1. This
table shows that most supersonic events occur within granular cells or
at the edges of granules (70\% of the total). A small percentage
appear in intergranular lanes (6\%), whereas 7\% occur in what we call
\emph{evolving granulation}: a granule that turns into an intergranular
lane as the event takes place or vice-versa. A non-negligible amount
(9\%) could not be classified since they occurred at locations where
the LOS velocities are very close to zero. Finally, 8\% of all events
were classified as \emph{exotic}. By exotic we refer to those events
that occur in tiny granules that are surrounded by large downflows.

Figures 3--7 illustrate some examples of the above events. In these
figures, the first row corresponds to the continuum intensity at each
pixel normalized to the average quiet Sun continuum intensity, $I_{\rm
c}/I_{\rm qs}$.  The second row shows the circular polarization in the
continuum: $V_{\rm c}/I_{\rm qs}$. This is the parameter that has been
used to identify the supersonic patches (see Section 3): $| V_{\rm c}
|/I_{\rm qs} > 0.015$. Regions that satisfy this condition are
enclosed in all panels by the black contours. The third row displays
the LOS velocity.  Note that the LOS velocity was calculated from a
Gaussian fit to Stokes $I$, and therefore these velocities are not
necessarily equivalent to the velocities deduced from $V_{\rm c}$
(Section 3). The forth row shows the line-averaged linear polarization
($\sqrt{Q^2+U^2}$). By line-average we refer to the sum over the first
four filter positions (see yellow squares in Figure 1). Finally, the
fifth row displays the line-averaged circular polarization. In this
case, the line-average sums the absolute value of the circular
polarization at each of the first four filter positions. The sign of
the line-averaged circular polarization is then taken from the sign of
Stokes $V$ at the first filter position. With this convention,
positive values of the line-averaged circular polarization denote a
magnetic field that points upwards from the solar surface or in order
words, $\gamma < 90^{\circ}$ (with $\gamma$ being the inclination of
the magnetic field vector with respect to the vertical direction on
the solar surface\footnote{In fact $\gamma$ denotes the inclination of
the magnetic field with respect to the observer's LOS, but at disk
center (where these observations where taken) it coincides with the
vertical direction on the solar surface}). Negative values of the
line-averaged circular polarization indicate exactly the opposite,
that is, a magnetic field pointing downwards into the Sun ($\gamma >
90^{\circ}$).

Figure 3 highlights the most common type of supersonic event: one
occurring within a granular cell in a {\it single} patch of enhanced
$V_{\rm c}$. Because it appears inside a granule, the associated
velocities are always blueshifted.  This indicates, as discussed in
Section 3, that the observed signal in $V_{\rm c}$ is not due to
\ion{Fe}{i}~5250.217~{\AA} but to \ion{Fe}{i}~ 5250.653~{\AA}. This
kind of event was extensively analyzed in \cite{2010ApJ...723L.144B}, where it
was associated with the presence, in most cases, of opposite
polarities in the magnetic field and significant linear polarization
in the surroundings of the event. The example in Fig.~3 clearly fits
this description. This led us to conclude that these events occur as a
consequence of magnetic reconnection.

Figure 4 displays an example of a \emph{double} event, where two
patches of opposite $V_{\rm c}$ appear next to each other.  The upper
one has $V_{\rm c}<0$, while the lower one possesses $V_{\rm
c}>0$. The fact that in both regions the velocities are blueshifted
indicates that the different sign in $V_{\rm c}$ cannot be produced by
one patch harboring an upflow and the other a downflow. Instead, the
reason for the two different signs is to be found in the inclination
of the magnetic field. This is demonstrated by the fifth row in
Fig.~4, where the line-averaged Stokes $V$ signal has a different
sign. This configuration can be best explained by the lower patch
being in a environment with $\gamma < 90^{\circ}$ (magnetic field
pointing upwards), while the upper patch has $\gamma > 90^{\circ}$
(magnetic field points downwards). Attending to the line-averaged
linear polarization (fourth row) the two polarities are connected by a
horizontal magnetic field. This example is consistent with the
configuration of a magnetic loop that reconnects.  Therefore, this
situation corresponds to a very similar case as in Fig.\ 3 with the
exception that each polarity (footpoints of the loop) develops a
supersonic magnetic upflow.

\begin{figure*}[p]
\begin{center}
\includegraphics[width=11.9cm,bb=54 395 479 1068,clip]{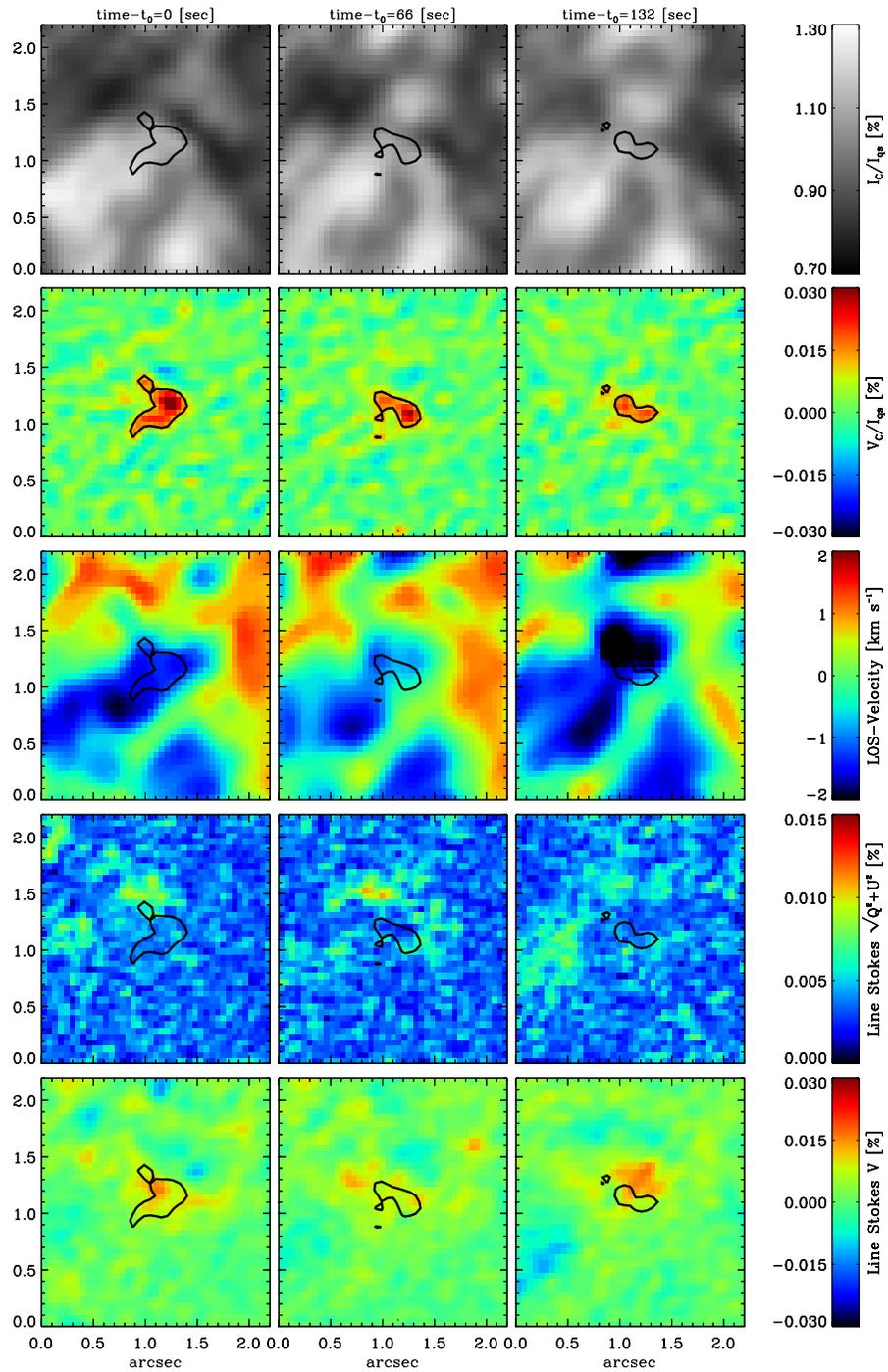}
\caption{Example of a supersonic event classified as {\it single} 
and occurring within a granular cell. From top to bottom: continuum
intensity, circular polarization at the continuum, LOS velocity, and
line-averaged linear and circular polarization. {\em From left to right:}
snapshots at different times during the evolution of the event.}
\end{center}
\end{figure*}

\begin{figure*}[p]
\begin{center}
\includegraphics[width=11.9cm,bb=54 395 479 1068,clip]{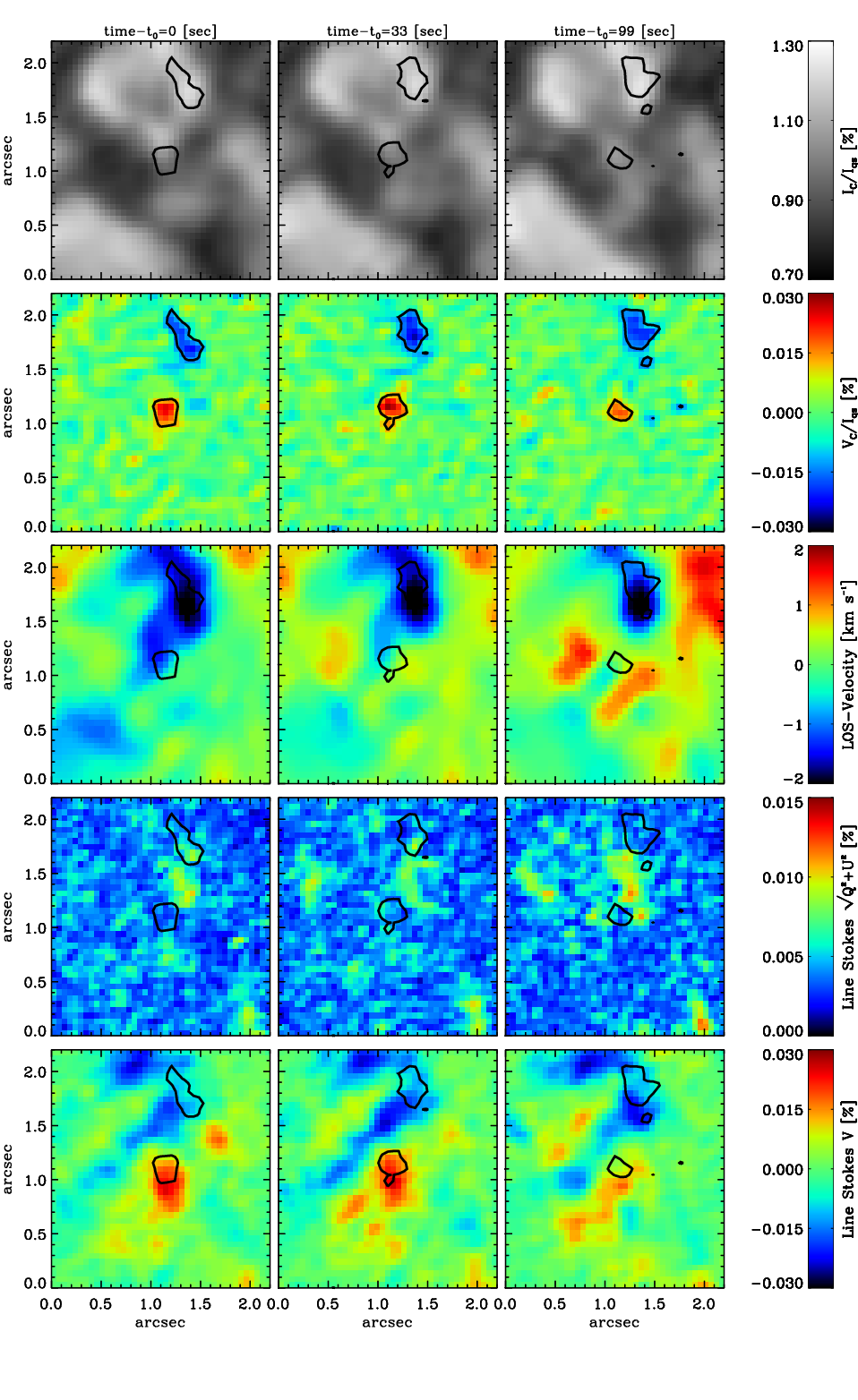}
\caption{Same as Fig.~3 but for a supersonic event 
classified as {\it double}. See text for details.}
\end{center}
\end{figure*}

Figure 5 shows an example of a \emph{twin} event, in which two patches
of $V_{\rm c}/I_{\rm qs} > 0.015$ of the same sign appear.  The upper
one occurs at the center/edge of a granular cell, whereas the lower
develops around a region of \emph{evolving granulation}: an initial
downflow that becomes an upflow later on. The upper patch coincides
with a region of positive polarity and therefore the observed $V_{\rm
c}$ signal corresponds to a blueshift from
\ion{Fe}{i}~5250.653~{\AA}. The lower patch is more difficult to
identify, as the polarity of the magnetic field is not clear and the
associated velocities change sign in time. Nevertheless, the presence
of magnetic fields of opposite polarities in a 2\arcsec\/ area around
the event is evident. As in Fig.\ 4, the two patches of enhanced
$V_{\rm c}$ are connected by horizontal magnetic fields (as seen in
the linear polarization panels). Again, this magnetic configuration
suggests the presence of a magnetic loop undergoing a reconnection
event.

Figure 6 illustrates an example of an \emph{exotic} event. This
particular case has been classified as such because of the unusual
velocity pattern. This pattern reveals that the supersonic events
occur at the edge of a tiny granule surrounded by very strong
downflowing lanes. The presence of opposite polarities in the
line-averaged Stokes $V$ signal (fifth panel) reveals the presence of
opposite polarities in the surroundings, which again points towards a
reconnection event. The line-averaged linear polarization (fourth
panel) does not show particularly large horizontal fields, thus making
it difficult to associate this event to a magnetic loop.

The final case that we will discuss in this paper is an example of a
supersonic event taking place in a
\emph{downflow} lane (Figure 7). The amount of $V_{\rm c}$ observed at the event location denotes that the downflow 
is magnetized. This signal is however negative, i.e., $V_{\rm c} <
0$. This indicates that the red lobe of Stokes $V$ from
\ion{Fe}{i}~5250.217~{\AA} shifts into the continuum wavelength
($\Delta\lambda=227$ m{\AA}) due to a strong downflow embedded in a
positive polarity magnetic field ($\gamma < 90^{\circ}$). The
supersonic event occurs at the center of the downflowing lane, at a
location where an enhancement of the continuum intensity is also seen
(upper panel).  Remarkably, no clear opposite polarities are detected
in the surrounding FOV during the event, and no large linear
polarization is observed either. Therefore this event does not fit
with the aforementioned scenario of reconnection of magnetic
loops. Instead, it resembles more closely the initial stages of
\emph{convective collapse}
\citep{2001ApJ...560.1010B}.  However, this interpretation is not completely
clear either, as the lack of large total (circular plus linear)
polarization signals is not compatible with the presence of strong
(kG) magnetic fields \citep{2008ApJ...677L.145N,2009A&A...504..583F}.  A possible
solution is that the strong downflow shifts the Stokes profiles away
from the scanned wavelength region, thereby giving the impression that
the magnetic field is low.

\begin{figure*}
\begin{center}
\includegraphics[width=11.9cm,bb=54 395 479 1068,clip]{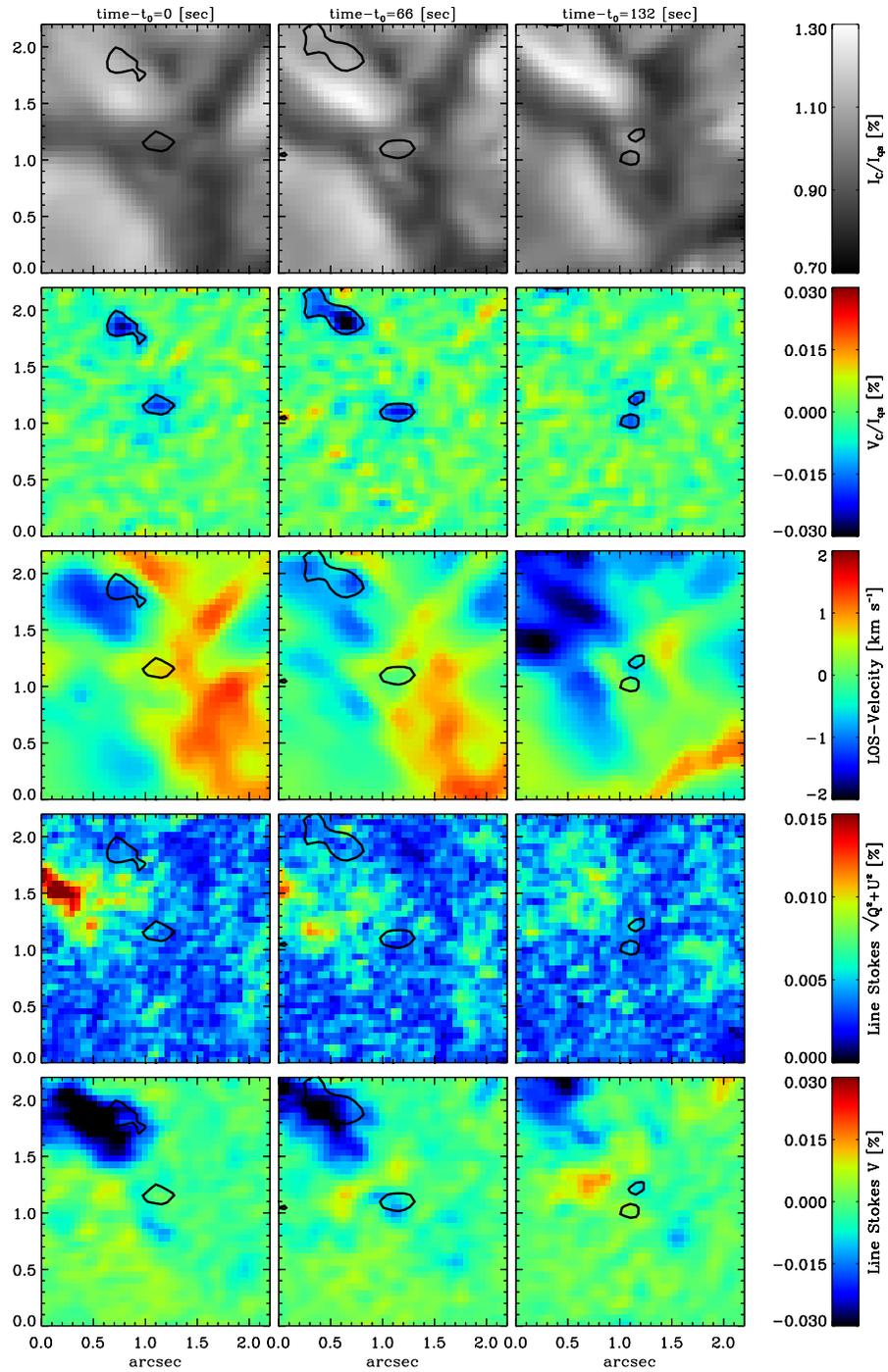}
\caption{Same as Fig.~3 but for a supersonic event classified as {\it twin}. See text for details.}
\end{center}
\end{figure*}

\begin{figure*}
\begin{center}
\includegraphics[width=11.9cm,bb=54 395 479 1068,clip]{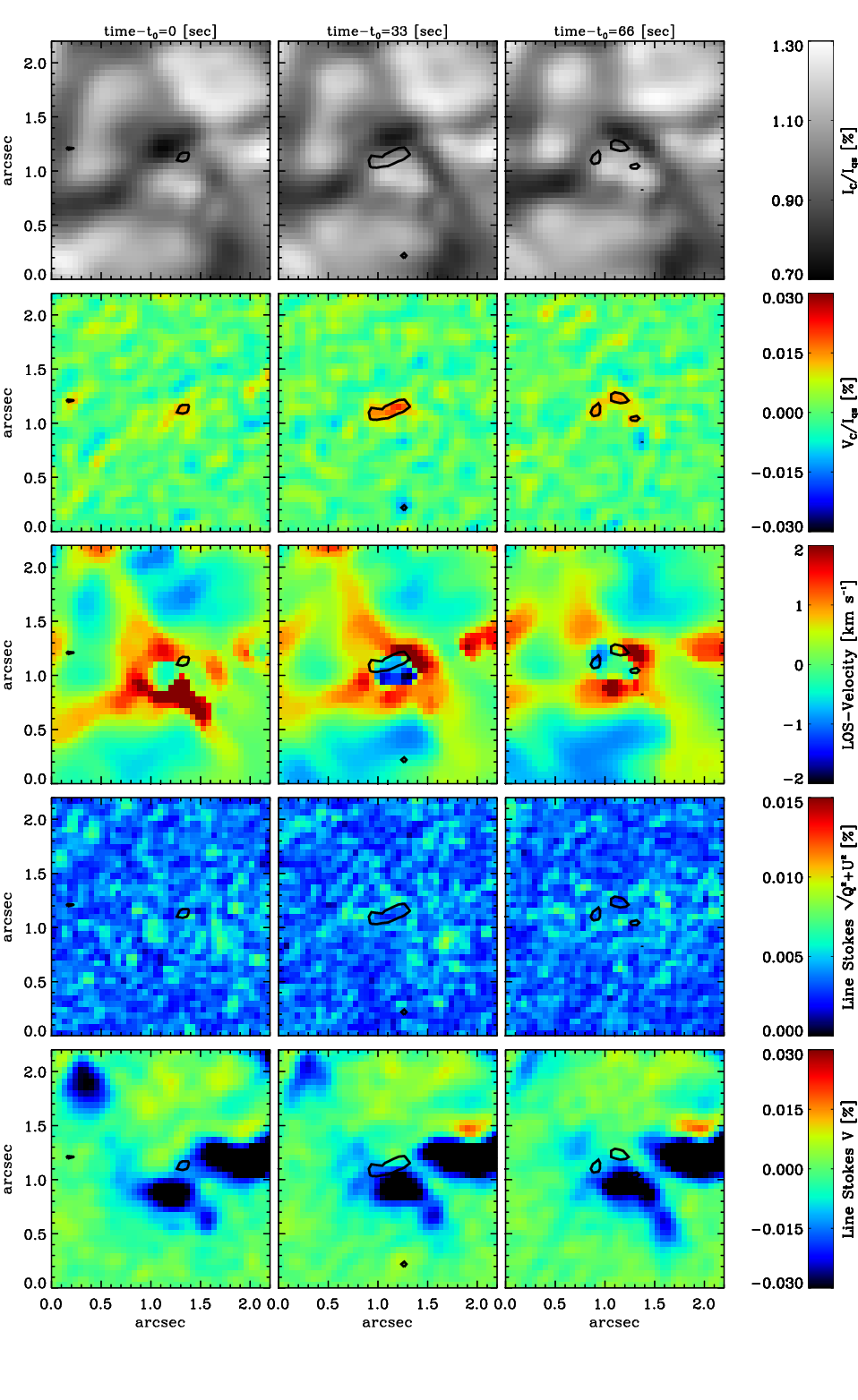}
\caption{Same as Fig.~3 but for a supersonic event classified as {\it exotic}. See text for details.}
\end{center}
\end{figure*}

\begin{figure*}
\begin{center}
\includegraphics[width=11.9cm,bb=54 395 479 1068,clip]{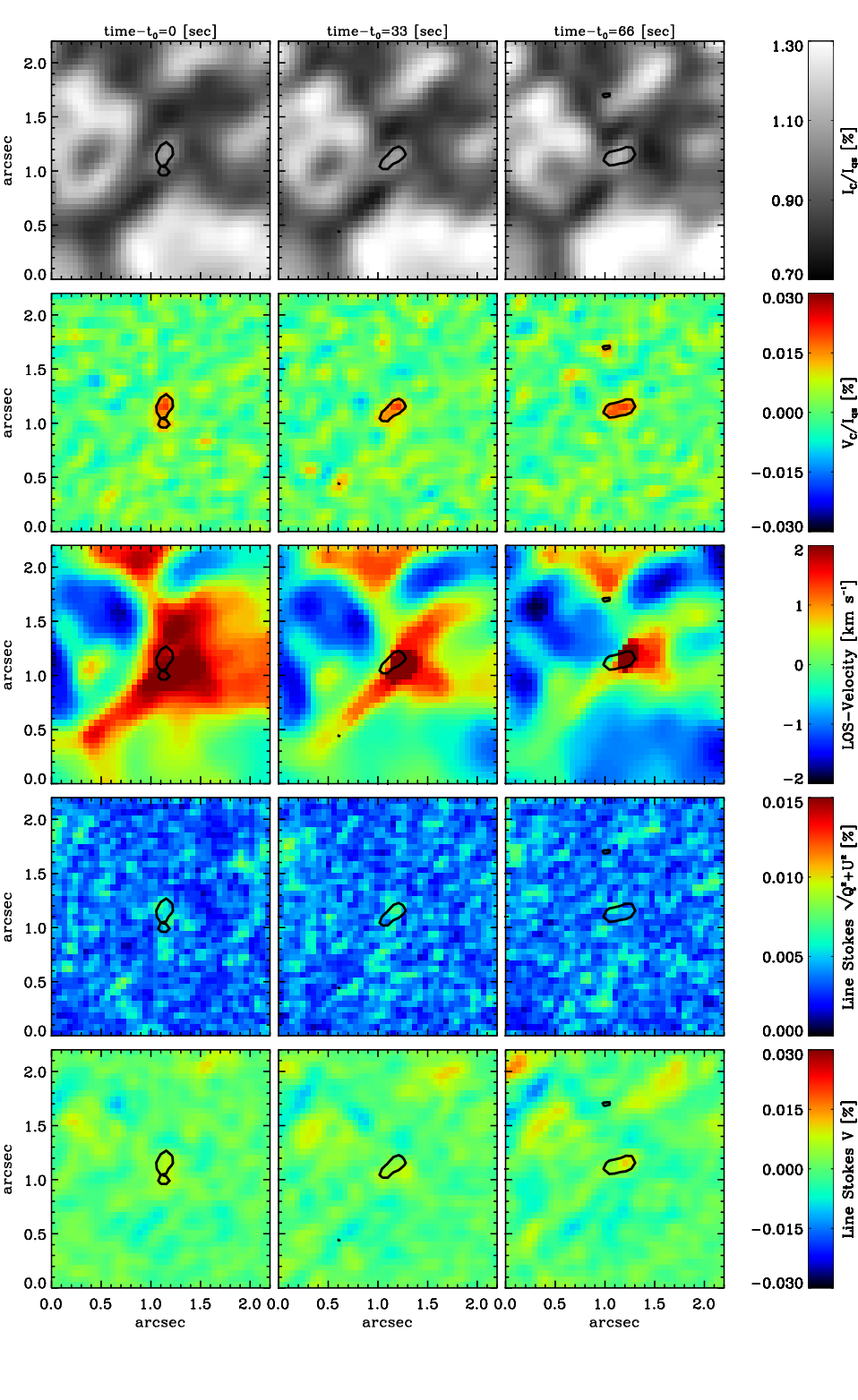}
\caption{Same as Fig.~3 but for a supersonic event occurring in a downflowing lane. See text for details.}
\end{center}
\end{figure*}

\section{Conclusions}

In this paper we have presented some of the recent observations
carried out by the IMaX instrument on-board the stratospheric balloon
{\sc Sunrise}. These observations comprise full Stokes polarimetry of
the quiet Sun with a high temporal cadence and extremely high spatial
resolution. This unique dataset has allowed us to detect a number of
supersonic magnetized flows in the quiet Sun that appear as highly
shifted circular polarization signals. We have illustrated several of
these events and we have been able to interpret them in terms of
events associated with the emergence of magnetic loops, magnetic
reconnection, and convective collapse.

Because IMaX observes only 5 wavelength positions around \ion{Fe}{i}
5250.217~{\AA}, it is important to confirm our findings with
additional observations. A first attempt has already been made by
\cite{2011A&A...530A.111M} who, employing data from the {\em Hinode} 
spectro-polarimeter, have also detected these supersonic events in the
quiet Sun. Their main result is that they seem to appear indistinctly
in granular and intergranular lanes. In the future, hopefully some of
these events will also be observed with instruments with better
spectral and spatial resolution, thereby helping to narrow down the
list of possible physical processes responsible for them. Instruments
in future and current large solar telescopes such as NST
\citep{2010AN....331..620G}, GREGOR \citep{2010AN....331..624V}, ATST 
\citep{2010AN....331..609K},
and EST \citep{2010SPIE.7733E..15C} should also contribute to study
these phenomena in more detail.

\bibliography{hinode4}

\end{document}